\documentclass[aps,prb,twocolumn,showpacs,preprintnumbers,amsmath,amssymb,epsfig]{revtex4}

\usepackage{graphicx}
\usepackage{graphics}
\usepackage{epsfig}
\usepackage{dcolumn}
\usepackage{bm}
\newcounter{tr}
\newcommand{\bqe}{\begin{eqnarray}}
\newcommand{\eqe}{\end{eqnarray}}

\begin{document}
\title{ Theory of Brillouin Light Scattering from Ferromagnetic Nanospheres}
\author{Ping Chu}
 \email{pingc@uci.edu}
 \affiliation{Department of Physics and Astronomy, University of
California, Irvine CA 92697, USA}

\author{D. L. Mills}
 \email{dlmills@uci.edu}
 \affiliation{Department of Physics and Astronomy, University
of California, Irvine CA 92697, USA}

\date{\today}

\begin{abstract}
We develop the theory of Brillouin light scattering (BLS) from spin
wave modes in ferromagnetic nanospheres, within a framework that
incorporates the spatial variation of the optical fields within the
sphere. Our recent theory of exchange dipole spin wave modes of the
sphere provides us with eigenvectors. When properly normalized,
these eigenvectors allow calculation of the absolute cross section
of various modes which contribute to BLS spectrum. We then present
explicit calculation of the BLS spectrum associated with the first
few dipole/exchange spin wave modes with emphasis on their relative
intensity.
\end{abstract}
 \maketitle
\section{Introduction}

As lithographic techniques have rapidly developed in the last
decade, it has become possible to assemble ordered magnetic elements
on the nanometer scale. These have attracted interest as potential
candidates in magnetic data-storage devices. Spin wave excitations
control the dynamic response of the magnetization in the linear
response regime, and the speed of real devices at least for small
amplitude motions of the magnetization. This provides motivation for
theoretical and experimental studies of spin waves in magnetic
wires, dots and other nanoscale structures\cite{2,3,4,5}.

In our previous paper\cite{1}, we provided an analytical method to
calculate eigenvalues and eigenvectors of dipole-exchange spin waves
in ferromagnetic nanospheres, and we explored the properties of
these modes. A question is how one may access these modes
experimentally. We argue here that Brillouin light scattering
spectroscopy (BLS) offers the means of probing the dipole-exchange
spin waves of very small spheres. In this paper, we develop the
theory of BLS from a single metallic nanosphere, and illustrate with
explicit calculations that one may access the spectrum of
dipole/exchange spin waves modes by this means.

As we know, there are two complementary experimental techniques to
study spin waves in small structures such as nanoscale particles or
spheres: ferromagnetic resonance (FMR) spectroscopy and Brillouin
light scattering (BLS). FMR uses microwaves to excite spin waves,
but the microwave wavelength is much larger than the radii of
nanospheres. It is the case as well that their radius is far smaller
than the skin depth of microwaves, which is typically a micron for
materials of current interest. Thus, FMR produces a uniform exciting
field inside nanospheres. This makes it impossible to excite the
full spectrum of dipole-exchange modes of such an object. Only the
uniform mode is excited in this circumstance. However, the laser
photon used in BLS has a small skin depth at optical frequencies.
This is typically 10-20 $nm$ in the ferromagnetic metals of
interest. This is comparable with the radius of nanoscale spheres,
and the resulting non-uniform optical field can thus, in principle,
excite a full spectrum of dipole-exchange modes. It is then of
interest to develop an explicit description of the scattering cross
sections associated with the various modes, and to obtain insight
into the factors which control these cross sections. This is the
purpose of the present paper. We shall see below from the
calculations presented in the present paper that can indeed excite
and study higher order dipole exchange modes through the BLS method.
We shall conclude that BLS is a powerful tool which one can provide
information about the nature of the spin wave spectrum of nanoscale
spheres. This will be true of samples of less simple shape as well,
of course. A quantitative theory of the BLS spectrum of more complex
objects is quite non trivial, unfortunately. While our calculations
apply only to small spheres, our principal conclusion that the
method can excite the dipole-exchange spin wave spectrum of
nanoscale magnetic entities is quite general, in our view.

From the quantum point of view, the process of BLS in ferromagnets
is viewed as an interaction between light photons and magnons. From
the classical point of view, the thermal fluctuations of
magnetization inside the sphere will change the dielectric tensor of
the ferromagnetic material in a time dependent manner. The inelastic
scattering event has its origin in the scattering of the laser light
from these dynamic fluctuations in the dielectric constant. We use a
semi-classic method here to address this issue.

In this paper, we present the theory of BLS from spin waves in
ferromagnetic nanospheres, and we calculate the cross section of BLS
from several low lying dipole exchange modes. We shall see that for
small spheres, the cross section for exciting particular higher
order dipole-exchange modes is appreciable, indeed larger than that
of the uniform mode for sufficiently small spheres.

\section{The Structure of the Theory}

We consider a metallic ferromagnetic nanosphere with radius in the
range of 10$nm$ to 60$nm$, with vacuum outside. Since the linear
dimensions of the sphere are large compared to the lattice constant,
this allows us to use continuum theory to describe the spin motions
in the sphere. The ferromagnetic nanosphere is magnetized uniformly
along z-direction with saturation magnetization $M_s$. At small but
finite temperatures, the thermally excited magnetization can be
written by $\vec{M}(\vec{r},t)=M_s\hat{z}+\vec{m}(\vec{r},t)$,where
$\vec{m}(\vec{r},t)$ is a small amplitude disturbance associated
with thermal spin waves. As noted earlier, the eigenvectors which
describe the dipole-exchange spin waves of the sphere can be
obtained from our previous paper\cite{1}. These allow us to describe
the spin wave contribution to $\vec{m}(\vec{r},t)$ provided that the
eigenvectors are properly normalized. We remark that the
prescription for normalizing such eigenvectors has been discussed
only quite recently\cite{6}.In the regime where dipolar
contributions to the excitation energy are appreciable, the
prescription for normalizing the eigenvectors is not so obvious.

The fluctuation of magnetization causes a time-dependent change of
the dielectric tensor of the material, which we write as $\delta
\epsilon_{\mu \nu} (\vec{r},t) = \sum \limits_\lambda
K_{\mu\nu\lambda} m_\lambda(\vec{r},t)$. We retain terms only first
order in the transverse magnetization, to generate a description of
BLS by processes in which a single spin wave quantum is created or
destroyed. When applied to static magnetization arrangements, the
term just described controls Faraday rotation. An early theoretical
analysis of the spin wave BLS spectra of 3d ferromagnetic films
shows that a fully quantitative account of surface and bulk spin
wave contributions follow by choosing
$K_{\mu\nu\lambda}=K\varepsilon_{\mu\nu\lambda}$, where
$\varepsilon_{\mu\nu\lambda}$ is the Levi Civita tensor  \cite{7},
appropriate to bulk materials of cubic crystal structure. We adopt
this picture here, so that we have
\bqe\delta\epsilon_{\mu\nu}(\vec{r},t)=K\sum\limits_\lambda\varepsilon
_{\mu\nu\lambda}m_\lambda(\vec{r},t).\label{eq:1}\eqe

As just discussed, thermal fluctuations produce time and space
varying components to the magnetization, which in turn lead to
fluctuations of the dielectric tensor. Now we turn to the
description of the light scattering of electromagnetic radiation by
dielectric tensor fluctuations in the sphere. We describe the
electromagnetic fields in the vicinity of the sphere by Maxwell's
equations: \bqe
\nabla\times\vec{E}=-\frac{1}{c}\frac{\partial\vec{B}}{\partial
t},\label{eq:2} \eqe \bqe
\nabla\times\vec{B}=\frac{1}{c}\frac{\partial\vec{D}}{\partial
t},\label{eq:3} \eqe where within the sphere, \bqe
D_\mu(\vec{r},t)=\epsilon E_\mu(\vec{r},t)+\sum\limits_\nu \delta
\epsilon_{\mu\nu}(\vec{r},t) E_\nu(\vec{r},t), \label{eq:4} \eqe
with $\epsilon$ is the complex dielectric constant of the material
from which the sphere is fabricated. Since $\delta
\epsilon_{\mu\nu}(\vec{r},t)$ varies slowly in time with respect to
the laser field, from Eq.(\ref{eq:4}) we have, \bqe \frac{\partial
D_\mu (\vec{r},t)}{\partial t}=-i \epsilon \omega E_\mu
(\vec{r},t)-i \omega\sum\limits_\nu \delta\epsilon_{\nu}(\vec{r},t)
E_\nu(\vec{r},t),\label{eq:5}\eqe where $\omega$ is the frequency of
the incident laser. Upon inserting this into Maxwell's equations, we
have \bqe \nabla\times\vec{E}(\vec{r},t)=i \epsilon \frac{\omega}{c}
\vec{B}(\vec{r},t),\label{eq:6} \eqe \bqe
\nabla\times\vec{B}(\vec{r},t)=-i \epsilon \frac{\omega}{c} \vec{E}
(\vec{r},t)+\frac{4\pi}{c}\vec{J}^{eff} (\vec{r},t).\label{eq:7}
\eqe We have defined $J_\mu^{eff}(\vec{r},t) =\frac{\omega}{4\pi i}
\sum\limits_\nu \delta \epsilon_{\mu\nu}(\vec{r},t) E_\nu
(\vec{r},t)$. Upon using the relation between the dielectric tensor
and the magnetization as in Eq.(\ref{eq:1}), we obtain the form of
the effective current density, given by \bqe
J_\mu^{eff}(\vec{r},t)=\frac{\omega K}{4\pi
i}\sum\limits_{\nu,\lambda} \varepsilon_{\mu\nu\lambda}
m_\lambda(\vec{r},t)E_\nu(\vec{r},t).\label{eq:8}\eqe Since the
thermal fluctuations are very small in amplitude, we will treat the
effect of the term involving $\vec{J}^{eff}$ in Eq.(\ref{eq:7}) in
the Born approximation. Thus, we may replace $E_\mu(\vec{r},t)$ by
$E_\mu ^{(0)}(\vec{r},t)$ in Eq.(\ref{eq:8}), with $E_\mu ^{(0)}$
the amplitude of the incident field. Thus, we have \bqe
J_\mu^{eff}(\vec{r},t)=\frac{\omega K}{4\pi
i}\sum\limits_{\nu,\lambda} \varepsilon_{\mu\nu\lambda}
m_\lambda(\vec{r},t)E_\nu ^{(0)}(\vec{r},t).\label{eq:9}\eqe

The problem is now that of computing the radiation fields generated
by the effective current density within the sphere. Of interest to
us are radiation fields in the far zone, as $r\rightarrow\infty$. We
may proceed by following the approach outlined by Jackson \cite{8}.
Far from the sphere, the radiation fields may be described as a
superposition of fields of TM and TE character. In the far zone, the
TM component of the radiation field has the magnetic field
perpendicular to $\hat{r}$, the unit vector in the radial direction,
while the electric field has both a radial and transverse component.
The TE mode has the character of $\vec{E}$ and $\vec{B}$
interchanged. Given the radial component of $\vec{E}$ in the far
zone, one can derive expressions for the transverse components of
both $\vec{E}$ and $\vec{B}$ for the TM mode. Furthermore, one can
derive equations satisfied by $\vec{r}\cdot \vec{E}$ everywhere:
\begin{subequations}
\bqe [\nabla^2+\frac{\omega^2}{c^2}\epsilon(\omega)] (\vec{r}\cdot
\vec{E}) =-\frac{4\pi\omega}{i c^2} \vec{r}\cdot\vec{J}^{eff},
(r<R),\label{eq:10a} \eqe \bqe [\nabla^2+\frac{\omega^2}{c^2}]
(\vec{r}\cdot \vec{E}) =0 , (r>R).\label{eq:10b} \eqe
\end{subequations}
Similar statements can be made regarding the TE fields. One can
derive equations satisfied by $\vec{r}\cdot\vec{B}$, and given
$\vec{B}$ in the far zone, one can derive the remaining components
of $\vec{B}$. The combination $\vec{r}\cdot\vec{B}$ satisfies:
\begin{subequations}
\bqe [\nabla^2+\frac{\omega^2}{c^2}\epsilon(\omega)] (\vec{r}\cdot
\vec{B}) =\frac{4\pi\omega}{i c^2} \vec{L}\cdot\vec{J}^{eff}
,(r<R),\label{eq:11a} \eqe \bqe [\nabla^2+\frac{\omega^2}{c^2}]
(\vec{r}\cdot \vec{B}) =0 , (r>R),\label{eq:11b} \eqe where
$\vec{L}=\frac{1}{i}\vec{r}\times\vec{\nabla}$ is the orbital
angular-momentum operator of quantum mechanics.
\end{subequations}

We can calculate the radial components of the electromagnetic fields
by using Green's functions. By having solutions for the radial
components of $\vec{B}$ and $\vec{E}$, we can generate a complete
description of the TM and TE multipole mode fields in the far zone,
where the fields have radiative character. The analysis proceeds
very much along the lines given in Jackson's text\cite{8}, so we
only provide the final results. The transverse field components in
the far zone are of interest, since these control the radial flow of
outgoing scattered energy. The expressions for these field
components in the far zone may be written
\begin{subequations}
\begin{widetext}
\bqe E_\theta ^{(tot)}=i\frac{e^{ik_0
r}}{r}\sum\limits_{l,m}\frac{1}{(i)^l l (l+1)} [\Lambda_l ^{(TM)}
\Gamma_{l,m} ^{(TM)} L_\varphi Y_{l,m} (\theta,\varphi)-\Lambda_l
^{(TE)} \Gamma_{l,m} ^{(TE)} L_\theta Y_{l,m}
(\theta,\varphi)],\label{eq:12a}\eqe \bqe E_\varphi
^{(tot)}=-i\frac{e^{ik_0 r}}{r}\sum\limits_{l,m}\frac{1}{(i)^l l
(l+1)} [\Lambda_l ^{(TM)} \Gamma_{l,m} ^{(TM)} L_\theta Y_{l,m}
(\theta,\varphi)+ \Lambda_l ^{(TE)} \Gamma_{l,m} ^{(TE)} L_\varphi
Y_{l,m} (\theta,\varphi)],\label{eq:12b}\eqe \bqe B_\theta
^{(tot)}=i\frac{e^{ik_0 r}}{r}\sum\limits_{l,m}\frac{1}{(i)^l l
(l+1)} [\Lambda_l ^{(TM)} \Gamma_{l,m} ^{(TM)} L_\theta Y_{l,m}
(\theta,\varphi)+ \Lambda_l ^{(TE)} \Gamma_{l,m} ^{(TE)} L_\varphi
Y_{l,m} (\theta,\varphi)],\label{eq:12c}\eqe \bqe B_\varphi
^{(tot)}=i\frac{e^{ik_0 r}}{r}\sum\limits_{l,m}\frac{1}{(i)^l l
(l+1)} [\Lambda_l ^{(TM)} \Gamma_{l,m} ^{(TM)} L_\varphi Y_{l,m}
(\theta,\varphi)- \Lambda_l ^{(TE)} \Gamma_{l,m} ^{(TE)} L_\theta
Y_{l,m} (\theta,\varphi)],\label{eq:12d}\eqe
\end{widetext}
\end{subequations}where $k_0=\omega/c$, and we note the angular momentum operators
$L_\varphi=-i\frac{\partial}{\partial \theta}$ and
$L_\theta=\frac{i}{\sin\theta}\frac{\partial}{\partial\varphi}$. The
quantities $\Gamma_{l,m} ^{(TM)}$ and $\Gamma_{l,m} ^{(TE)}$ can be
viewed as the $(l,m)$ components of the effective current densities,
with origin in the thermal fluctuations of the magnetization, as we
have seen. The source $\Gamma_{l,m} ^{(TM)}$ generates scattered
radiation of TM character, while $\Gamma_{l,m} ^{(TE)}$ generates
scattered radiation of TE character. We have the forms
\begin{subequations}
\bqe \Gamma_{l,m} ^{(TM)}=\int \frac{d^3 r}{R} j_ { l} (k r) Y_{l,m}
^\ast(\theta,\varphi)
\vec{r}\cdot\vec{J}^{eff}(\vec{r}),\label{eq:13a}\eqe and \bqe
\Gamma_{l,m} ^{(TE)}=\int d^3 r j_ { l} (k r)
\vec{L}\cdot\vec{J}^{eff}(\vec{r}) Y_{l,m}
^\ast(\theta,\varphi),\label{eq:13b}\eqe  where $k=k_0
\sqrt{\epsilon} $ is the (complex) wave vector of electromagnetic
radiation inside the sphere. The imaginary part of $k$ controls the
penetration depth of the incident laser radiation.
\end{subequations}

In Eq.(12), the factors $\Lambda_l ^{(TM)}$ and $\Lambda_l ^{(TE)}$
have the following physical interpretation. The effective currents
just described create scattered radiation inside the nanosphere;
this radiation must be transmitted through the surface of the sphere
to the outside. These quantities are the transmission coefficients
for TM and TE waves of partial wave character $l$, respectively. We
need to discuss properties of these transmission coefficients to
appreciate the reason why the numerical results in section III
display the systematics we shall see. We begin by quoting the
general expressions for $\Lambda_l ^{(TM)}$ and $\Lambda_l ^{(TE)}$,
and then we pause to explore their behavior in the limit of interest
to us.

We express the factors as
\begin{subequations}
\begin{widetext}
\bqe \Lambda_l ^{(TM)}=\frac{4 \pi k_0 k^2 \epsilon R^2}{c}
(\frac{j_{ l} (k R)h_{ l-1} ^{(1)}(k R)-j_{ l-1}(k R) h_{ l}
^{(1)}(k R)}{\epsilon H_l(k_0 R)j_{ l}(k R)-h_{ l}^{(1)}(k_0 R)G_l(k
R)}),\label{eq:14a}\eqe \bqe &\text{and        }& \Lambda_l
^{(TE)}=\frac{4 \pi k^2 R}{c}(\frac{j_{ l} (k R)h_{ l-1} ^{(1)}(k
R)-j_{ l-1}(k R) h_{ l} ^{(1)}(k R)}{H_l(k_0 R)j_{ l}(k R)-h_{
l}^{(1)}(k_0 R)G_l(k R)})\label{eq:14b}.\eqe
\end{widetext}
\end{subequations}

We have introduced $G_l(x)=x j_{ l-1} (x)-l j_{ l} (x)$ and $H_l
(x)=x h_{l-1}^{(1)} (x)-l h_l^{(1)} (x)$, where $j_{ l} (x)$ is the
spherical Bessel function of the first kind, and $h_l^{(1)} (x)$ is
the spherical Hankel function of the first kind. The wave vector $k$
in Eq.(14) is again the (complex) wave vector of electromagnetic
radiation within the material from which the nanosphere is
fabricated. Thus, the magnitude of the product $kR$ is roughly the
radius of the nanosphere divided by the optical skin depth $\delta$.
As noted in section I, this ratio will be in the range of unity, for
samples of interest. The quantity $k_0$ is the free space wave
vector of electromagnetic radiation of frequency $\omega$ and for
spheres whose diameter is near the smaller end of the the range of
10-60 $nm$ we shall have $k_0 R\ll 1$. Thus, of interest is the
relative magnitudes of $\Lambda_l ^{(TM)}$ and $\Lambda_l ^{(TE)}$
when $k_0 R\ll 1$. Upon using small argument forms for the relevant
Hankel functions, one finds in this limit

\begin{subequations}
\begin{widetext}
\bqe \Lambda_l ^{(TM)}\cong\frac{(k_0 R)^{l+2}}{(2l-1)!!}\{\frac{4
\pi k^2 R}{c}(\frac{j_{ l} (k R)h_{ l-1} ^{(1)}(k R)-j_{ l-1}(k R)
h_{ l} ^{(1)}(k R)}{l j_{ l}(k R)+G_l(k
R)/\epsilon})\},\label{eq:15a}\eqe \bqe &\text{and        }&
\Lambda_l ^{(TE)}\cong\frac{(k_0 R)^{l+1}}{(2l-1)!!}\{\frac{4 \pi
k^2 R}{c}(\frac{j_{ l} (k R)h_{ l-1} ^{(1)}(k R)-j_{ l-1}(k R) h_{
l} ^{(1)}(k R)}{l j_{ l}(k R)+G_l(k R)})\}. \label{eq:15b}\eqe
\end{widetext}
\end{subequations}In Eq.(15a) and Eq.(15b), the factors in curly
brackets are close in magnitude to each other. We thus see that in
the limit that the wavelength of the laser radiation is large
compared to the radius of the sphere, the intensity of TE radiation
with angular momentum $l$ ( its strength is controlled
by$\Lambda_l^{(TE)}$) will be stronger than the TM radiation with
angular momentum $l$ ( its strength is controlled
by$\Lambda_l^{(TM)}$) by the large factor of $(k_0 R)^{-2}$. This
follows since $\Gamma_{l,m} ^{(TM)}$ and $\Gamma_{l,m} ^{(TE)}$ are
roughly the same order of magnitude. This observation will be very
important when we discuss the relative scattering intensities of the
various dipole-exchange spin wave modes below, in the limit the
radius of the sphere is small.

We have one final step. In the definition of the effective current
density $\vec{J}^{eff}(\vec{r},t)$ given in Eq.(\ref{eq:9}), we
require an expression for the incident laser field
$\vec{E}^{(0)}(\vec{r})$ inside the nanosphere. If we consider a
plane wave incident on a conducting sphere of arbitrary radius, the
expression for this field is cumbersome. Here we exploit the fact
that our sphere has radius $R$ small compared to the wavelength of
the laser light in vacuum. In this limit, to excellent
approximation, we may assume that the sphere is placed in the
spatially uniform field $\vec{E}^{(0)}=\hat{z} E^{(0)} exp(-i\omega
t)$. In the region just outside the sphere, we are far into the near
zone, to use language appropriate to radiation theory, so the
spatial form in the field outside the sphere may be taken to be that
given by the equations of electrostatics. However, the radius of the
sphere is not small compared to the optical skin depth, so we must
solve for the electric field inside through use of the full
electromagnetic theory. We surely have $\vec{\nabla}\cdot\vec{E}=0$
inside the sphere, so the electric field obeys the Helmholtz
equation \bqe \nabla^2 \vec{E} (\vec{r})+k^2 \vec{E}
(\vec{r})=0\label{eq:16}\eqe everywhere inside the sphere, where as
above $k^2=(\omega/c)^2 \epsilon$ with $\epsilon$ the optical
frequency dielectric constant. Outside the sphere, in response to
the incident field we have the classical electrostatic field of a
polarized sphere. In the near zone and this has components

\begin{subequations}
\bqe E_r ^>=\frac{2a}{r^3}\cos\theta  &\text{  } ; &\text{ }E_\theta
^>=\frac{a}{r^3}\sin\theta, \label{eq:17a}\eqe whereas inside the
sphere the fields have the form \bqe E_r ^<=b\frac{2 j_1 (k
r)}{r}\cos\theta ; \nonumber\\E_\theta ^<=-\frac{b}{r}\frac{d}{dr}
(r j_1 (k r))\sin\theta. \label{eq:17b}\eqe
\end{subequations}Application of the boundary conditions is straightforward, and yields
\bqe b=\frac{3}{2} \frac{E^{(0)} R}{\epsilon j_1 (k R)+\frac{d}{dr}
(r j_1(k R))|_R}. \label{eq:18}\eqe The expressions for optical
fields inside a conducting nanosphere under conditions similar to
those considered here will prove useful for a variety of
applications. Thus we pause by arranging the results into a more
useful format. We define
\begin{subequations}
\begin{widetext}
\bqe \bar{f}=\frac{R}{r}\frac{\epsilon +2}{(\epsilon -1) j_1 (k
R)+\sin (k R)}(j_1 (k r)+\sin (k r))\label{eq:19a}\eqe \bqe &\text {
and  }& \Delta f=\frac{R}{r}\frac{\epsilon +2}{(\epsilon -1) j_1 (k
R)+\sin (k R)}(3 j_1 (k r)-\sin (k r)). \label{eq:19b}\eqe
\end{widetext}
\end{subequations} Of course, inside the sphere, the field is
invariant under rotation around the z axis. The two non zero components
of the field in cylindrical coordinates are given by
\begin{subequations}
\bqe E_z (r)=\frac{3 E^{(0)}}{\epsilon +2} (\bar{f}+\Delta f (r)
\cos (2\theta)) \label{eq:20a}\eqe \bqe &\text{and        }& E_\rho
(r)=\frac{3 E^{(0)}}{\epsilon +2} \Delta f (r) \sin (2\theta).
\label{eq:20b}\eqe
\end{subequations} In the limit that the skin depth is large compared
 to the radius of the sphere, the function $\bar{f} (r)$  approaches
 unity everywhere, and $\Delta f (r)$ approaches zero. We are then left
 with the elementary expression for the fields inside a dielectric sphere
 exposed to a spatially uniform electric field. When the skin depth is
 comparable to or smaller than the radius of the sphere, the use of Eqs.(20)
 allows us to account for the spatially non uniform character of the laser
 field inside the sphere. It should be remarked that in the numerical
 calculations we report below, the magnetization of the sphere is always
 taken parallel to the $z$ axis, and we will explore the Brillouin cross
 section when the laser field is not necessarily parallel to the $z$ axis.
 Of course, in this case one may apply a simple coordinate rotation to the
 field components in Eqs.(20) to generate expressions for the exciting field
 inside the nanosphere.

We now have all the ingredients in place to obtain an expression for
the Brillouin scattering cross section. One evaluates the Poynting
vector in the far zone, integrates it over solid angle, and divides
the result by the energy per unit time which illuminates the sphere.
One finds the following for the total Brillouin scattering cross
section: \begin{widetext}\bqe
\sigma_{BLS}=\frac{1}{|E^{(0)}|^2}\sum_{l,m}\frac{1}{l
(l+1)}(|\Lambda_{l,m}^{(TM)}\Gamma_{l,m}^{(TM)}|^2+
|\Lambda_{l,m}^{(TE)}\Gamma_{l,m}^{(TE)}|^2).
\label{eq:21}\eqe\end{widetext} The expression in Eq.(21), written
out in full, is a quadratic form in the transverse magnetization
components, with terms proportional to the combination $m_\mu
(\vec{r},t) m_\nu (\vec{r} ',t)$ integrated over the volume of the
sphere; one has an integral over both $\vec{r}$ and $\vec{r} '$. One
takes a thermal average over this quadratic form. The resulting
expression provides one with the Brillouin cross section, integrated
over all solid angle, and integrated also over all spin wave modes
which contribute to the Brillouin signal. One can break the
correlation functions $<m_\mu (\vec{r},t) m_\nu (\vec{r} ',t)>$ down
into sums over each of the spin wave eigenmodes of the sphere, and
thus extract from Eq.(21) the Brillouin cross section integrated
over solid angle for each individual mode. When we refer to the
Brillouin cross section of an individual mode, we include both the
Stokes and the anti Stokes component. The algebra associated with
the decomposition just described is complex, and since the resulting
forms are of little general interest, we omit the details here. We
refer the reader to the discussion given in ref.[6] for a
description of how this is done. It is the case, as noted in this
paper, that in the regime where dipolar contributions to the spin
wave excitation energy are of comparable magnitude, the algorithm
for normalizing the eigenvectors is not so obvious, as noted above.
The means of doing so are derived in ref.[6].

We turn next to our numerical studies of the Brillouin scattering
intensities of the various normal modes.

\section{Results and Discussion}

We first begin with some general comments. It is the case that the
nature of the spin wave modes in ferromagnetic spheres and related
sample shapes is a classic topic in magnetism. Motivation for the
early theoretical studies was provided by the first generation of
FMR experiments, which were largely carried out on spherical samples
of yttrium iron garnet (YIG). These samples had linear dimensions
that were macroscopic. As a consequence, the long wavelength spin
wave modes excited by the microwave fields used in FMR could be
described by magnetostatic theory with exchange ignored. In a
remarkable paper that is now classic, Walker presented an analytic
theory of the magnetostatic spin wave modes of uniformly magnetized
ferromagnetic ellipsoids\cite{9}. The special case of the sphere was
discussed by Fletcher and Bell, who provide a very large and
interesting array of useful formulae for descrbing the
eigenfrequencies and eigenvectors as well\cite{10}. As the radius of
the sphere is made small, to the point where the radius lies in the
few nanometer range, exchange enters importantly in the description
of the spin wave eigenmodes. This is particularly the case for the
materials of current interest, which are the 3d ferromagnets or
alloys fabricated from these elements. By virtue of the itinerant
character of their ferromagnetism, along with their high Curie
temperatures, the exchange stiffness $D$ is very large. The first
description of the exchange/dipole spin wave modes of a
ferromagnetic sphere has been developed only recently by Arias and
the present authors\cite{1}; this paper extends the analysis of
ref.[9] and ref.[10] to include exchange as well as dipolar
interactions between the spins. A complete discussion of the
response of such a sphere to spatially inhomogeneous microwave
frequency magnetic fields is given as well.

As discussed in ref.[1], a measure of the importance of exchange in
the description of the spin wave modes of a ferromagnetic sphere is
the dimensionless parameter $r_{ex}=\frac{D}{4\pi
M_s}(\frac{\pi}{R})^2$, where $R$ is the radius of the sphere.
Consultation of Fig.4 of ref.[1] shows that the contribution of
exchange to the excitation energy of spin waves in spheres is
substantial even for values of $r_{ex}$ as small as 0.2.  The
calculations presented below are for $Fe$ nanospheres, for which $4
\pi M_s=2\times 10^4 gauss$, and $D=2.5\times 10^{-9} gauss\cdot
cm^2$. For these parameters, and for an Fe sphere with radius of 10
$nm$, we have $r_{ex}\cong 1$. We call the reader's attention to the
Appendix of ref.[1] where simple analytic expressions are provided
for the excitation energies of spin waves in the limit of strong
exchange, with the first correction from dipolar interactions
included. As noted in ref.[1], this expression accounts nicely for
the excitation energies calculated from the full theory for a rather
wide range of $r_{ex}$.

In this paper, we refer to the dipole exchange spin wave modes by
the quantum number scheme used in ref.[1], which is adapted from
ref.[9] and ref.[10].  First, suppose we consider the magnetostatic
limit. Then for each eigenmode of the ferromagnetic sphere, the
magnetic dipole field generated outside the sphere by the spin
motion has an angular dependence controlled by a single spherical
harmonic $Y_{l,m} (\theta,\varphi)$. Inside the sphere, the magnetic
potential which describes these fields is not simple, if wishes to
express it in terms of spherical harmonics. It is the case that each
mode can be labeled by the two quantum numbers $l,m$ associated with
the spherical harmonic just mentioned, and a radial quantum number
$n$ which may be viewed as a radial quantum number, to make an
analogy with quantum mechanics. Hence, in the magnetostatic limit,
each mode is labeled by the three numbers $(l,m,n)$. The uniform
mode, which is the mode seen in FMR when the sphere is exposed to a
spatially uniform microwave field, is the (1,1,0) mode in this
notation. When exchange is added, it is no longer true that the
magnetic potential outside the sphere is proportional to a single
spherical harmonic. However, as the limit that the exchange
stiffness $D$ is allowed to approach zero, each mode of the
exchange/dipole spectrum smoothly and continuously reduces to a
particular magnetostatic mode which can be labeled by the scheme
just described. Thus, we label each dipole/exchange mode of the
sphere by the same three quantum numbers employed for the mode in
the limit $D\rightarrow 0$. We recall ref.[1] that with no spin
pinning at the surface of the sphere (the case considered here), the
FMR mode is unaffected by exchange, and remains a uniform mode even
with exchange present. We now turn to our results.

\begin{figure}

\includegraphics[scale=.5]{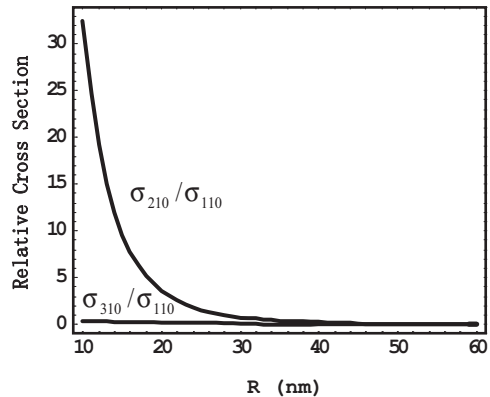}
    \caption{For the case where the electric field of the laser is
    parallel to the magnetization, we plot the cross section ratios
    $\sigma_{210}/\sigma_{110}$ and $\sigma_{310}/\sigma_{110}$ as
    a function of the radius of an Fe nanosphere. The influence of exchange,
     as a function of radius, for these and other dipole/exchange modes is given
     in Fig.4 of ref.[1].}
    \label{fig1}
\end{figure}

\begin{figure}

\includegraphics[scale=.5]{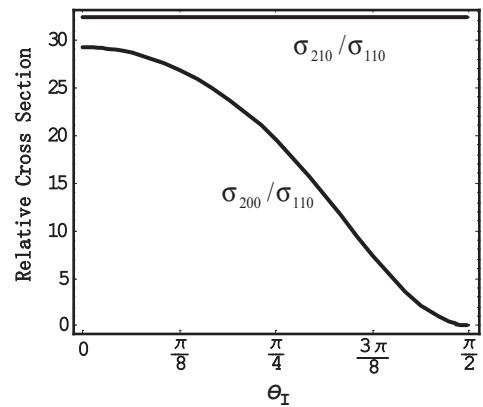}
    \caption{We show the dependence of the cross section ratios
    $\sigma_{210}/\sigma_{110}$ and $\sigma_{200}/\sigma_{110}$ on the angle $\theta_I$
     between the magnetization of the sphere and the electric field of the incident laser beam.
      The calculations are for an Fe sphere whose radius is 10 $nm$.}
    \label{fig2}
\end{figure}

We have carried out calculations of the BLS cross sections for the
(110) mode, the (2m0) modes and the (3m0) modes, and also the (410)
mode. Since there is little interest in the absolute BLS cross
section for a single, isolated sphere, we shall present the result
in the form of cross section ratios of the form
$\sigma_{lm0}/\sigma_{110}$. Our principle conclusions are
summarized in Fig.1 and Fig.2, for the case where the incident
photons have the energy of 2.5 eV. In Fig.1, for the case where the
electric field vector of the incident radiation  is aligned with the
magnetization of the sphere, we show two of the cross section ratios
as a function of the radius of the nanosphere. We have taken
material parameters appropriate to Fe, but clearly the trends are
robust, and not affected by details of the material parameters. We
see that as the radius of the sphere is reduced, the scattering
power of the (210) mode increases dramatically relative to the FMR
mode, while that of the (310) mode remains roughly similar to that
of the FMR mode.

What is responsible for the behavior of the cross sections in Fig.1
is the operation of what we may call a quasi selection rule. First,
for the FMR mode, one may show that all of the coefficients
$\Gamma_{l,m}^{(TE)}$ vanish identically, and the only non zero
contribution to the cross section comes from $\Gamma_{l,m}^{(TM)}$.
Thus, we can say that in the limit when $k_0 R\ll 1$, the dominant
contribution to the BLS cross section for the FMR mode (that from
$\Gamma_{11}^{(TE)}$) is "silent", and the FMR mode scatters only by
virtue of the contribution $\Gamma_{11}^{(TM)}$, which as we see
from Eqs.(15) leads to a contribution to the cross section smaller
than that from the leading term by the factor of $(k_0 R)^2$. For a
nanosphere with a radius of 10 $nm$, we have $k_0 R\approx 0.13$,
whereas when we reach 50 $nm$ radii, $k_0 R$ is sufficiently large
that the quasi selection rule is not operative. For the (210) mode,
$\Gamma_{11}^{(TE)}$ is substantial for the smaller spheres. The
quasi selection again suppresses the cross section of the (310) mode
in the limit of small radii.

\begin{table*}
\begin{tabular}{|c|c|c|c|c|c|}
\hline
Cross Section Ratio& $R=10 nm$ & $R=50 nm$ & Cross Section Ratio& $R=10 nm$ & $R=50 nm$ \\
at $\theta_I=0$ & & &at $\theta_I=\pi/2$ &{ }&{} \\
\hline
$\sigma_{200}/\sigma_{110}$&29.20&0.09&$\sigma_{200}/\sigma_{110}$&0.09&0.001\\
\hline
$\sigma_{210}/\sigma_{110}$&32.43&0.03&$\sigma_{210}/\sigma_{110}$&32.43&0.02\\
\hline
$\sigma_{220}/\sigma_{110}$&0&0.01&$\sigma_{220}/\sigma_{110}$&0&0.005\\
\hline
$\sigma_{300}/\sigma_{110}$&1.13&0.49&$\sigma_{300}/\sigma_{110}$&0&0\\
\hline
$\sigma_{310}/\sigma_{110}$&0.37&0.07&$\sigma_{310}/\sigma_{110}$&1.51&1.70\\
\hline
$\sigma_{320}/\sigma_{110}$&0.01&0.001&$\sigma_{320}/\sigma_{110}$&0.007&0.001\\
\hline
$\sigma_{330}/\sigma_{110}$&0&0&$\sigma_{330}/\sigma_{110}$&0&0\\
\hline
$\sigma_{410}/\sigma_{110}$&0&0.02&$\sigma_{410}/\sigma_{110}$&0&0.01\\
\hline
\end{tabular}
\caption{For spheres with radii of 10 $nm$ and 50 $nm$,
respectively, we tabulate the cross section ratios
$\sigma_{lm0}/\sigma_{110}$ for the various dipole/exchange modes
considered. We provide results for the case where the incident laser
field is parallel to the magnetization ($\theta_I=0$) and the case
where it is perpendicular ($\theta_I=\pi/2$). Zeros are inserted
into the table for modes with very small excitation cross sections.}
\end{table*}

In Table I, we tabulate the cross section ratio for several modes,
and for sphere radii of 10 $nm$, and also 50 $nm$. We see that for
the larger sphere, the (110) has the largest cross section, but the
(310) mode is nearly as strong. As the sphere is reduced in size the
quasi selection rule dominates, and the $l=2$ modes dominate the
spectrum, to have very large cross sections save for the $m = 2$
case, where the cross section is very small; we insert a zero into
the Table for this reason.

The angular dependence of the cross section ratios is also of
interest. In Fig.2, we show the cross section ratios as a function
of the angle between the incident laser field and the magnetization
for the two modes which dominate the spectrum for small spheres. We
see that in the limit $k_0 R\ll 1$ the large cross section for the
(210) mode remains large for all angles, whereas the scattering from
the (200) mode is suppressed as electric field of the incident laser
is rotated from parallel to the magnetization, to the equatorial
plane of the sphere. Experimental observation of this angular
dependence would be of great interest, as a test of the picture put
forward in this paper.

Insight into the origin of the quasi selection rule which controls
the results above in the limit of small radii can be obtained from
the limiting forms displayed in Eqs.(15) and in the structure of the
quantities $\Gamma_{l,m}^{(TE)}$ and $\Gamma_{l,m}^{(TM)}$ defined
in Eqs.(13). For the form of the coupling between the light and the
spin system described earlier,  and for the case where the incident
electric field is aligned with the magnetization of the sphere, one
has the explicit forms:
\begin{subequations}
\begin{widetext}
\bqe \Gamma_{l,m}^{(TE)}=A\int_V d^3 r j_{ l} (k r)
\bar{f}(r)[\frac{\partial Y_{lm} (\theta,\varphi)^*}{\partial
\theta} m_\rho (\vec{r})+i m \cot (\theta) Y_{lm} (\theta,\varphi)^*
m_\varphi (\vec{r})] \label{eq:22a}\eqe  \bqe &\text{and  }&
\Gamma_{l,m}^{(TM)}=i A \int_V \frac{d^3 r}{R} r j_l (k r)
\bar{f}(r) Y_{lm} (\theta,\varphi)^* \sin (\theta) m_\varphi
(\vec{r}). \label{eq:22b}\eqe \end{widetext}
\end{subequations} In these expressions, the integration is over the
volume of the nanosphere, we have defined $A=[3 \omega K
E^{(0)}/4\pi(\epsilon+2)]$, and the components of the transverse
magnetization associated with a given spin wave mode are expressed
in cylindrical coordinates.

From the structure of the eigenvectors given in ref.[1], one may see
that each of the dipole/exchange spin wave modes of the sphere has
well defined parity, even or odd, under reflection through the
equatorial plane of the sphere. The parity of the mode may be
written as $(-1)^{l+m}$. Now the spherical harmonic which appears in
Eq.(22b) along with the factor of $\sin(\theta)$ also have parity
$(-1)^{l+m}$ under this reflection, whereas the parity of the
integrand in Eq.(22a) is $(-1)^{l+m+1}$. Then we also note that for
a given mode with label $(l_0,m_0,n)$ the angular integration over
$\varphi$ renders all the quantities in Eqs.(22) zero except for the
contributions for which $m=m_0$.

Thus, we can see from the previous paragraph that the term which
provides the leading contribution to the cross section,
$\Gamma_{11}^{(TE)}$ vanishes for the (110) mode, which is the mode
excited in FMR. It is the case that for this mode, the only non zero
contribution to these coefficients is $\Gamma_{11}^{(TM)}$. The FMR
mode is thus forbidden to scatter in lowest order, in the limit $k_0
R\ll 1$.  When we turn to the $l=2$ modes, for the (200) mode
$\Gamma_{10}^{(TE)}$ is non zero and substantial, and for the (210)
mode the same is true of $\Gamma_{11}^{(TE)}$. For the (220) mode,
the azimuthal integrations render all the leading terms,
$\{\Gamma_{1,m}^{(TE)}\}$, zero and we obtain our first substantial
contribution from $\Gamma_{22}^{(TM)}$. The cross section for this
mode is thus very small, in the limit $k_0 R\ll 1$. For the (310)
mode, $\Gamma_{11}^{(TE)}$ is forbidden by reflection symmetry
through the equatorial plane, and the first non zero contribution is
from $\Gamma_{11}^{(TM)}$. The cross section for this mode is thus
comparable to that for the (110) mode, the FMR mode. We can continue
on to understand the small cross sections for the (320) and the
(330) mode. By the time we reach the (410) mode, the angular
momentum content of the spin wave is sufficiently high that the
scattering cross section is small.

From the calculations presented in this section, we can see that BLS
spectroscopy can allow one to access selected higher order dipole
exchange modes, whereas ferromagnetic resonance spectroscopy probes
only the single FMR mode in the limit of small sphere radii.

\section{Concluding Remarks}

From the calculations presented here, we can see that BLS
spectroscopy of the spin wave modes of small nanospheres can provide
one with access to certain higher order dipole/exchange modes. In
the limit of small sphere radii, in the sense described above, we
see that the $l=2$ modes are predicted to dominate the spectrum.

The analysis presented here explores the BLS cross section for the
various spin wave modes, for a single, isolated nanosphere. Of
course, any experimental study will necessarily explore an array of
such objects. Indeed, Steinmuller et al. have reported BLS studies
of Fe nanospheres deposited on a GaAs surface\cite{11}. It is
difficult to comment in detail on their results, since the film
consisted of clusters of Fe spheres arranged in a disordered fashion
on the surface. We can comment on one issue, however. These authors
observe Stokes/anti Stokes asymmetries in their spectra, and argue
this provides evidence that the spin excitations they study should
be regarded as collective excitations of the array of spheres, which
interact through dipolar coupling. We recall that Stokes/anti Stokes
asymmetries in BLS spectra taken from the surfaces of ferromagnetic
crystals and in films were reported many years ago, and theory based
on the physical picture used here provides a fully quantitative
account of the data\cite{12,13,14}. A question raised with these
experiments is whether the Stokes/anti Stokes asymmetry is truly
evidence for collective mode behavior, or whether in fact such
asymmetries are found for the scattering of light from spin waves in
a single, isolated sphere such as that studied here. In section III,
we presented only total cross sections, integrated over both the
Stokes and anti Stokes features in the spectrum. We remark that in
the course of this investigation, we have inquired into the possible
occurrence of Stokes/anti Stokes asymmetries for the single sphere,
to find the answer in the negative. Thus, we reinforce the
interpretation offered in ref.[11].

It would be highly desirable to see BLS studies from ordered
lattices of ferromagnetic nanospheres, as opposed to the disordered
films studies in ref.[11]. Then, of course, one would observe
collective spin wave modes of the lattice of spheres where the
intersphere interaction is dipolar in nature, if one assumes that
neighboring spheres do not touch. We remark that we have developed a
theoretical description of such collective modes, for the case where
both exchange and intrasphere dipolar interactions are comparable in
strength\cite{15}. In our paper, one finds explicit calculations
which describe the nature of such modes. It will be a
straightforward matter to adapt the description of the BLS spectrum
of an isolated sphere set forth in the present paper to the
description of scattering from the collective modes. Experimental
data on ordered arrays of spheres will provide motivation for this
extension.

\acknowledgments

    This research was supported by the U. S, army, through Contract No.
    CS000128. We have enjoyed discussion of Brillouin light
    scattering from nanospheres with Prof. J.A.C. Bland.

\end{document}